\documentclass[%
 reprint,
 amsmath,amssymb,
 aps,
]{revtex4-2}

\usepackage{float}
\usepackage{subfloat}
\usepackage{amsbsy}
\usepackage{amsfonts}
\usepackage{amssymb}
\usepackage{tabularx,ragged2e,booktabs}
\usepackage{amsmath}
\usepackage{breqn}
\usepackage{graphicx}
\usepackage{bm}
\usepackage{color}
\usepackage{dcolumn}
\usepackage[]{breqn}

    \usepackage{psfrag}
    \usepackage{lipsum}
    \usepackage{subfig}
    \usepackage{caption}
    \usepackage{lineno}

\begin{document}

\preprint{APS/123-QED}

\title{Direct Numerical Simulation of Vortex-Induced Instability Over a Semi-Infinite Flat Plate with Discrete Roughness}

\author{Aditi Sengupta}
\email{aditi@iitism.ac.in}
\affiliation{Mechanical Engineering$,$ IIT (ISM) Dhanbad$,$ Jharkhand 826004$,$ India}

\author{Vajjala K. Suman}
\email{vksuman@iitk.ac.in}
\affiliation{Scientist$,$ National Aerospace Laboratories Bangalore$,$ Karnataka-560017$,$ India}

\author{Tapan K. Sengupta \footnote{Corresponding author.}}
\email{tksengupta@iitism.ac.in}
\affiliation{Mechanical Engineering$,$ IIT (ISM) Dhanbad$,$ Jharkhand 826004$,$ India}

\date{\today}

\begin{abstract}
The receptivity of flow over a semi-infinite flat plate to a combination of deterministic free stream excitation and surface roughness, is investigated by solving the incompressible Navier-Stokes equation (NSE). The free stream excitation is in the form of a convecting vortex with fixed strength and sign, while the roughness element is a Gaussian bump.  Two-dimensional (2D) and three-dimensional (3D) direct numerical simulations (DNS) are performed to determine whether the nonmodal nonlinear nature of transition, previously established for the vortex-induced instability in [Sundaram et al., “Effects of free stream excitation on the boundary layer over a semi-infinite flat plate,” Phys. Fluids, 32, 094110 (2020))], holds in the presence of surface roughness. The study aims to capture the fundamental differences arising due to the addition of surface roughness using instantaneous flow features, vorticity dynamics and application of nonlinear theories of receptivity. Roughness accelerates the transition process significantly, while retaining the indicator of nonmodal response, i.e. the spatio-temporal wave front (STWF).  The inherently 2D nature of the vortex-induced instability evolves to three-dimensionality much faster in the presence of roughness, necessitating a 3D framework to capture the effects of roughness accurately. The cumulative response to the surface bump and convecting vortex can be separated into four stages: (i) generation of vortical disturbances at the roughness, (ii) a spatio-temporal amplification of the disturbance field, (iii) emergence and amplification of the STWF marked by sequence of unsteady separation bubbles, and (iv) emergence of turbulent flow due to merging of the  unsteady separation bubbles to turbulent spots.
\end{abstract}

\maketitle


\section{\label{sec1}Introduction}

The route of transition to turbulence is varied depending on the nature of the underlying disturbances and the equilibrium field. The flow past a semi-infinite flat plate can be experimentally and numerically investigated to accurately obtain the equilibrium flow. Disturbances can be imposed on this equilibrium field from inside \cite{puckert2018experiments, sundaram2019tollmien} or outside \cite{matsubara2001disturbance, jacobs2001simulations, sengupta2019direct, sengupta2020nonmodal, sengupta2023compressibility} the boundary layer to trigger instabilities which can be traced with high fidelity computations \cite{sengupta2013high}. Prediction of transition and its control is semi-empirical to an extent due to the the existence of multiple routes to transition and a lack of a sufficiently clear mechanistic understanding of the underlying three-dimensional processes. The classical route via the amplification of Tollmien-Schlichting (TS) waves is induced by monochromatic excitation inside the boundary layer, experimentally shown by Schubauer and Skramstad \cite{schubauer1947laminar}. Dietz \cite{dietz1999local} harmonically vibrated a ribbon in the free stream to create TS waves inside the boundary layer on a flat plate with roughness using monochromatic excitation. However, in the presence of acoustic or free stream excitation, the flow followed an alternative route, broadly classified as bypass transition \cite{saric2002boundary, morkovin1993bypass}. There have been numerous studies on the mechanism behind bypass transition viz. transient growth \cite{trefethen1993hydrodynamic}, nonmodal linear \cite{sundaram2020effects} and nonmodal nonlinear \cite{sengupta2020nonmodal, eckhardt2007turbulence, sengupta2020effects2} depending upon the nature of disturbance \cite{saric2002boundary}. In recent investigations \cite{sengupta2020nonmodal, sundaram2020effects}, it has been established that for the flow over a semi-infinite flat plate excited in the free stream, the route for transition to turbulence follows a nonmodal, nonlinear mechanism with the appearance and growth of the spatio-temporal wave-front (STWF) \cite{sundaram2019tollmien}. One of the objectives of the present work is to extend this receptivity analysis for a semi-infinite flat plate having a discrete roughness element in the form of a surface bump.

Studies devoted to the response due to free stream excitation \cite{jacobs2001simulations, sengupta2020effects} have often focused on the effect of free stream turbulence (FST), which consists of free stream disturbances with multiple time and spatial scales. This makes the realization of a receptivity study highly convoluted and difficult to interpret. For this reason, the unit process involved in the free-stream turbulence has been conceived as a line vortex of constant strength created in the free stream which is convecting with constant speed \cite{sengupta2003vortex, lim2004visual, sengupta2020numerical}. The ensuing vortex-induced instability has received much attention \cite{peridier1991vortex, kendall1987experimental, sengupta2003vortex, lim2004visual, sengupta2018enstrophy, sengupta2019direct, sengupta2019tracking, sengupta2020nonmodal, sundaram2020effects} due to the direct correlation between the cause and effect, making it an ideal \lq \lq kernel\rq \rq experiment for receptivity studies. This deterministic free stream excitation problem has been identified as one of the most important unsolved problems of fluid dynamics. It explains transition to turbulence caused by unsteady separation due to fluctuating pressure induced by action of convecting vortices \cite{sengupta2021role, sundaram2022non}.

Kendall \cite{kendall1987experimental} performed experiments to quantify the effect of traversing pressure field and identified a nondimensional convection speed of 0.3, to be the most receptive to flow disturbances. This has been experimentally \cite{lim2004visual} and numerically confirmed by performing two-dimensional (2D) \cite{sengupta2003vortex, sengupta2018enstrophy} and three-dimensional (3D) \cite{sengupta2019direct, sengupta2019tracking} direct numerical simulations (DNS) of vortex-induced instability. Here, for this same convection speed, we are interested in recording the additional effects due to a surface bump on the flat plate. The results of Kendall \cite{kendall1987experimental} are difficult to interpret due to ambiguity about the strength and sign of the vortices associated with the pressure pulses, apart from the inability to trace the height of the free stream disturbance field. This prompted a careful design of an experiment by Lim {\it et al.} \cite{sengupta2003vortex} which provided a direct causal relationship for vortex-induced receptivity. The authors created a captive vortex,
whose strength, sign, and height could be reproduced at will for the reported experiment. The experiment is driven by the free
stream vortex creating an adverse pressure gradient that destabilizes the boundary layer and leads to unsteady separation. Peridier {\it et al.} \cite{peridier1991vortex} discussed vortex-induced boundary layer separation caused by a single vortex released in quiescent ambiance. Interestingly, the authors noted numerical problems due to vortical eruptions which was considered as an indicator of transition to turbulence. Obabko and Cassel \cite{obabko2002navier}, take the Batchelor vortex as the primary vortex moving at a constant speed along the wall, as the model to study unsteady separation of flow. Two-dimensional simulations of the vortex-induced instability \cite{sengupta2003vortex} showed the bypass transition mechanism involved by investigating the sources/sinks of the disturbance mechanical energy (DME). Three-dimensional simulations investigated this using the disturbance enstrophy transport equation (DETE) \cite{sengupta2019direct} and showed the inherent 2D nature of the excitation, eventually leading to three-dimensionality due to the spatio-temporal growth of vortex-induced instability. The presence of STWF for flows excited from the free stream has been conclusively shown in 2D \cite{lim2004visual, sengupta2018enstrophy, sengupta2020nonmodal, sundaram2020effects} and 3D \cite{sengupta2019direct, sengupta2019tracking} simulations.

Surface roughness can have a profound effect on the transition process by inducing undesirable early transition of the boundary layer. Examples of roughness-induced transition include effect of discrete elements over aircraft wings, environmentally accumulated roughness through insect impacts or degradation of gas turbine components \cite{bons2010review}. There have been several DNS \cite{choudhari1993boundary, rizzetta2007direct, de2013laminar, suryanarayanan2019roughness, sengupta2017roughness} of roughness-induced transition, which support the view that roughness elements create distortions of the flow field displaying a nonmodal nature. Such disturbances are observed to be responsible for the eventual transition, thus this can be considered to be a case of bypass transition. Here, we intend to determine the mechanism followed by the transition to turbulence for a flow excited both from the free stream and on the surface in the form of bump. While there are numerous studies on the cumulative effects of FST and roughness, the effect of roughness in conjunction with the deterministic free stream excitation of the convecting vortex has not been analyzed before. This gap will be bridged here by performing a series of 2D and 3D DNS on a semi-infinite flat plate by solving the incompressible Navier-Stokes equations (NSE) using highly accurate numerical methods \cite{sengupta2013high}. Further analysis of flow features will be explored using nonlinear theories of receptivity based on DETE and DME. It will be determined whether the nonmodal nonlinear nature of transition previously established for the vortex-induced instability \cite{sengupta2020nonmodal, sengupta2019direct, sundaram2019tollmien} holds true in the presence of roughness.

The paper is organized as follows: In Sec. II, the 2D formulation of the vortex-induced instability problem with a surface bump is described. The governing equations and boundary conditions are explained and equilibrium flow is characterized. In Sec. III, the bypass transition process is compared for 2D vortex-induced instability with and without the roughness element using vorticity dynamics. The nonlinear theories of receptivity are described and applied to the 2D vortex-induced receptivity problem, and effect of including roughness is discussed in Sec. IV. In Sec. V, the 3D DNS for the receptivity problem with a Gaussian surface bump on the semi-infinite flat plate is formulated. The flow fields with and without roughness are compared using vorticity, disturbance enstrophy and DME. The paper closes with a summary and concluding remarks.

\section{Formulation of Vortex-Induced Instability with a 2D Surface Roughness Element}
\begin{figure*}
\begin{center}
\includegraphics[width=0.85\textwidth]{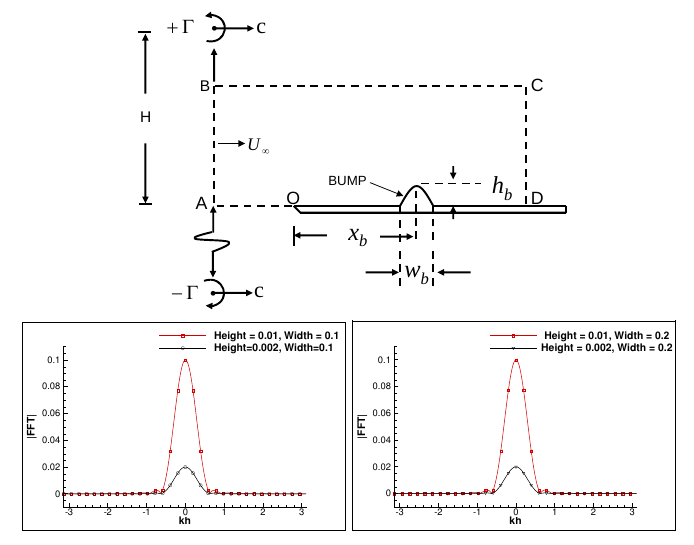}
\end{center}
\caption{Computational domain used in studying vortex-induced instability by an isolated vortex of circulation $\Gamma$, convecting at a speed $c$, at a constant height of $H$, in the free stream over a boundary layer formed over a semi-infinite flat plate with a surface bump. The spectrum associated with the Gaussian bump has also been shown, as function of wavenumber ($kh$).}
\label{fig1}
\end{figure*}

Vortex-induced instability is a typical example of bypass transition caused by a convecting vortex in the free stream. The problem has received attention in carefully designed receptivity experiments \cite{lim2004visual, kendall1987experimental} and in simulations \cite{sengupta2003vortex, sengupta2018enstrophy, sengupta2019direct, sengupta2020nonmodal, sundaram2020effects}. Here, we intend to show the mechanism of the vortex-induced instability triggered by the addition of a discrete roughness element on the semi-infinite flat plate considered in Fig. \ref{fig1}. In this section, we will show the formulation in the 2D framework by considering test cases with and without the roughness element. The inherently 2D nature of the disturbance field has been shown in prior studies on vortex-induced instability without the roughness element \cite{sengupta2019direct, sengupta2020nonmodal}.  For both the 2D and 3D simulations reported here, the same velocity-vorticity formulation has been used with identical numerical methods. Details of the numerical methods, staggered grids, interpolation schemes and grid generation have been provided in the appendix of Ref. [\onlinecite{sengupta2018enstrophy}] and the same is adopted in the present simulations.

Time-dependent boundary forcing by the convecting vortex excites the physical instability during the receptivity stage. Before we study the receptivity problem, the equilibrium flow is calculated by solving the NSE without the external forcing. The origin of the coordinate system is placed at the leading edge of the plate, whereas the computational domain starts upstream of it. In the simulations reported here, all velocities are scaled with the free-stream velocity, $U_{\infty}$, and all lengths with $L$, such that the Reynolds number is given by, $Re_L = U_{\infty} L/ \nu = 10^5$, where $\nu$ is the kinematic viscosity.

A potential line-vortex with counter-clockwise circulation ($\Gamma$), translates from left to right with a constant speed, $c$, at
a constant height, $H$, above the plate in Fig. \ref{fig1}. This vortex convecting in the inviscid part of the flow, imposes an image vortex below the plate, to ensure that the zero wall-normal velocity boundary condition is satisfied. A roughness element (Gaussian bump) is placed at a distance of $x_b = 1.5$ from the sharp leading edge of the semi-infinite flat plate for the cases with roughness. The height of the bump is $h_b$, while its width is $w_b$, as shown in Fig. \ref{fig1}. The equation for the Gaussian bump is given by,

$$y(x) = \frac{h_b}{2} \biggl[ 1 + \cos\biggl[\frac{2\pi(x - x_b)}{w_b}\biggr]\biggr],\; x_b - \frac{w_b}{2} \le x \le x_b + \frac{w_b}{2}$$
Characterizing the bump with $h_b$ and $w_b$ in the physical plane in terms of its slope (defined by the ratio of height to width of the bump), can alternately be presented in the wavenumber ($kh$) plane, as shown in the bottom frames of Fig. \ref{fig1}. The spectral amplitude is a strong function of $h_b$ when width is kept the same. For the same slope of the bump ($=0.1$), change of width from 0.1 to 0.2 does not change the bandwidth of scales excited by the bump, even though the spectral amplitude increases in proportion to the physical height.

\subsection{Velocity-Vorticity Formulation for 2D Flows}

The vorticity transport equation has been found to be accurate, as it is devoid of pressure-velocity coupling problem \cite{sengupta2013high}. In this formulation of the NSE, velocity field is obtained from the solution of the velocity Poisson equations. For the 2D problem, the velocity components, $u$, $v$, and the nonzero vorticity, $\omega$, are governed by equations given in Cartesian coordinates \cite{daube1992resolution, wu1995effective, joshi2023exploring}.

Here, the governing equations are solved in the transformed plane using highly accurate compact schemes to discretize the nonlinear convective acceleration terms. The governing vorticity transport and velocity Poisson equations in the transformed $(\xi, \eta)$-plane (with $\xi$ along streamwise and $\eta$ in the wall-normal directions) are given as

\begin{dmath}
\frac{\partial \omega}{\partial t} + \frac{1}{h_1} \frac{\partial }{\partial \xi}(u \omega) + \frac{1}{h_2}
\frac{\partial}{\partial \eta}(v \omega)  =  \frac{1}{Re} \frac{1}{h_1 h_2} \biggl[\frac{\partial}{\partial \xi}
\biggl(\frac{h_2}{h_1} \frac{\partial \omega}{\partial \xi}\biggr)
+ \frac{\partial}{\partial \eta} \biggl( \frac{h_1}{h_2} \frac{\partial \omega}{\partial \eta} \biggr)\biggl] + \nabla \times \vec{F}_i
\label{ch5_eq17}
\end{dmath}
\begin{eqnarray}
\biggl[ \frac{\partial}{\partial \xi} \biggl(\frac{h_2}{h_1} \frac{\partial u}{\partial \xi}\biggr) + \frac{\partial}{\partial
\eta} \biggl( \frac{h_1}{h_2} \frac{\partial u}{\partial \eta} \biggr)\biggl]  =  - \frac{\partial \omega}{\partial \eta}
\label{ch5_eq18}
\end{eqnarray}
\begin{eqnarray}
\biggl[ \frac{\partial}{\partial \xi} \biggl(\frac{h_2}{h_1} \frac{\partial v}{\partial \xi}\biggr) + \frac{\partial}{\partial
\eta} \biggl( \frac{h_1}{h_2} \frac{\partial v}{\partial \eta} \biggr)\biggl]  =  \frac{\partial \omega}{\partial \xi}
\label{ch5_eq19}
\end{eqnarray}

\noindent where $h_1$ and $h_2$ are the scale factors of the transformation given by, $h_1 = \sqrt{x_{\xi}^2+y_{\xi}^2}$ and
$h_2 = \sqrt{x_{\eta}^2+y_{\eta}^2}$. The curl of the body force appears on the right hand side of the vorticity transport equation (Eq. \eqref{ch5_eq17}). The divergence of the velocity field ($D_V = \nabla \cdot \overrightarrow{V} $) in the $(\xi, \eta)$-plane is given as

\begin{eqnarray}
D_V =\frac{1}{h_1 h_2} \biggl[\frac{\partial (h_1 v)}{\partial \eta} + \frac{\partial (h_2 u)}{\partial \xi}\biggl]
\label{ch5_eq20}
\end{eqnarray}

\noindent where divergence-free condition needs to be satisfied for mass conservation.

\subsection{Boundary Conditions}

The boundary conditions used in solving this problem are listed as follows:

(i) At the inflow (segment $AB$ of Fig.~\ref{fig1}), uniform inlet velocity $U_{\infty}$ is imposed, along with the contributions imposed by the free stream vortex and its image. Similarly, the boundary condition for $v$-velocity is calculated as caused by the convecting vortex and its image. The induced stream function created by the finite-core line vortex in the inviscid part of the flow is given as,

\begin{dmath*}
    \psi_{\infty} = U_{\infty} y - (U_{\infty} - c) \biggl[ \frac{(y-H)(d/2)^2}{\bar{x}^2 + (y -H)^2} + \frac{(y+H)(d/2)^2}{\bar{x}^2 + (y+H)^2} \biggr] \hspace{2mm} + \frac{\Gamma}{4\pi} Ln \frac{\bar{x}^2 + (y + H)^2}{\bar{x}^2 + (y - H)^2}
\end{dmath*}

In the above equation, the vortex of strength $\Gamma$, has a core diameter $d$ and convects with a speed $c$, at a constant height, $H$. The displacement effect of the vortex is fixed by the relative speed, $(U_{\infty} - c)$, and the circulation effect is given by the last term on the right hand side (RHS). The time dependence of the boundary condition is due to the translated coordinate, $\bar{x}$, which represents the current streamwise location of the convecting vortex.

(ii) At the wall (OD), no-slip condition is imposed on $u$ and $v$. The wall vorticity is computed from its definition: $\omega = \nabla \times \vec{V}$.

(iii) In the upstream segment $AO$ of Fig.~\ref{fig1}, $\omega = v = 0$, due to symmetry. The boundary condition for $u$ is fixed by the condition, $\partial u/\partial y=0$.

(iv) At the far-field boundary (BC): $\omega=0$ and ($u$, $v$) are calculated by the Biot-Savart law caused by the free stream vortex and its image.

(v) At the outflow (CD), $\omega$ is calculated using Sommerfeld boundary condition given by
$$\frac{\partial \omega}{\partial t} + U_c \frac{\partial \omega}{\partial x} = 0$$
\noindent where $U_c$ is chosen as $U_{\infty}$. The condition used on $u$ is, $\partial^2 u/\partial x^2=0$, whereas the condition for $v$ is given by, $\partial v/\partial x=\omega+\partial u/\partial y$.

The roughness element is introduced in the flow using an immersed boundary method which makes use of a volume of fluid approach \cite{yuan2014numerical}. The no-slip boundary conditions at rough surfaces are imposed using this method. The volume fraction occupied by the fluid $\phi_i(x,y)$ in each cell is calculated in the pre-processing step. A value of $\phi_i = 1$ is used for the grid points outside the roughness elements and $0< \phi_i <1$ on the physical boundary of the roughness element. A body force $F_i$ is obtained from the velocity predicted without the forcing term $\hat{u}_i$, as

\begin{equation}
F_i(x,y) = -(1-\phi_i(x,y)) \frac{\hat{u}_i(x,y)}{\Delta t}
\end{equation}

The curl of the body force, $\nabla \times F_i$ is added to the RHS of the vorticity transport equation given by Eq. \eqref{ch5_eq17} for all successive iterations. This enables one to account for the physical boundary of the roughness element while solving a simple mesh without the roughness element.

\subsection{Characterizing the Equilibrium Flow with and without the Surface Roughness}
Having obtained the NSE in velocity-vorticity formulation and the associated boundary conditions, one obtains the equilibrium flow for the cases with and without the bump. For the equilibrium flow, the NSE is computed without the convecting vortex in the free stream, by altering the boundary conditions appropriately. Equations (\ref{ch5_eq17}) to (\ref{ch5_eq20}) have been solved for the cases with and without the bump.

The computational domain in the streamwise direction is, $-0.05 \le x \le 20$, and includes the leading edge of the plate. The height of the computational domain is taken as $y_{max} = 0.75$, with the wall resolution given by $\Delta y = 3.688 \times 10^{-4}$, which is finer than the wall resolution taken in prior studies \cite{sengupta2003vortex, sengupta2018enstrophy}. The spacing in the wall-normal direction is stretched following a tangent hyperbolic distribution \cite{sengupta2016roles}, with the stretching parameter of $\beta_y = 2.0$. The maximum spacing is $5.186497 \times 10^{-3}$, by taking 301 points in $y$-direction. The first grid point in the streamwise direction has spacing of $\Delta x_i = 0.009106986$, with spacing reducing up to the leading edge of the plate following tangent hyperbolic distribution with a stretching parameter, $\beta_x = 2.0$. From the leading edge of the flat plate to a streamwise distance of $x = 5$, the grid is stretched with identical value of $\beta_x$. From $x = 5$ to the outflow of the computational domain ($x = 20$), a uniform grid is used with a spacing of $\Delta x_f = 0.021630828$. Altogether there are 1001 points in the streamwise direction. Reynolds number based on displacement thickness at the outflow is 2432.44 and this domain is five times longer than that taken in prior studies \cite{sengupta2003vortex, sengupta2018enstrophy}. The flow using $(\overrightarrow{V}, {\omega})$-formulation is computed with a time step of $\Delta t = 8 \times 10^{-5}$. For spatial discretization of convective terms, an optimized upwind compact scheme \cite{sagaut2023global} (OUCS3) is used and for time integration, an optimized three-stage Runge-Kutta (ORK3) method \cite{sengupta2017global} is used.

\begin{figure*}
\begin{center}
\includegraphics[width=0.95\textwidth]{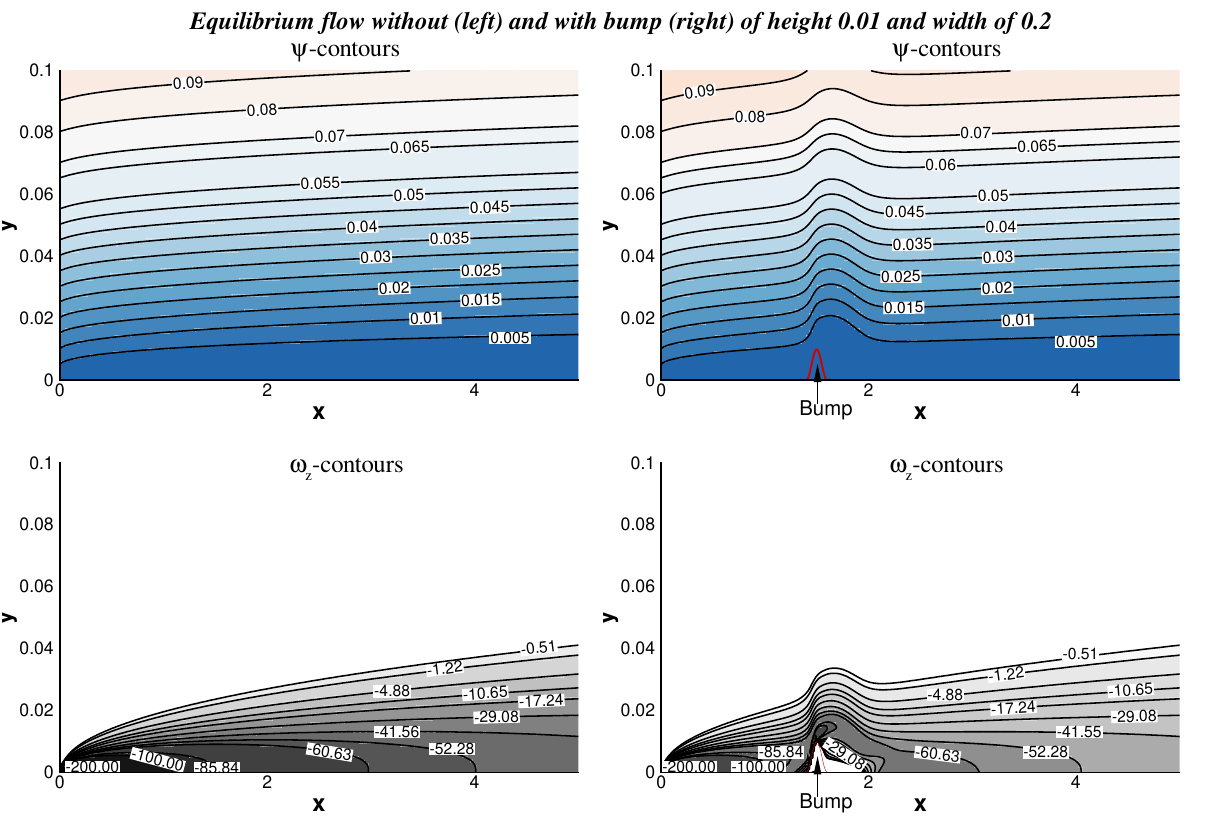}
\end{center}
\caption{Equilibrium flow solution for flow over a flat plate, with and without the surface bump. Stream function (top) and vorticity (bottom) contours are shown in the computational domain.}
\label{fig2}
\end{figure*}

In Fig.~\ref{fig2}, equilibrium flow for the cases of flow over semi-infinite flat plate with and without the bump is shown with the contour plots of stream function and vorticity. The time-dependent vorticity transport equation is solved till the maximum local acceleration in the full domain falls below the prescribed limit of $\frac{\partial \omega}{\partial t} \le 10^{-8}$. In the top right frame, for the case of equilibrium flow with bump, the flow immediately downstream of the bump is seen to be separated. This is the case of steady separation and is different from the unsteady separation noted for bypass transition \cite{lim2004visual, sengupta2003vortex, sengupta2020nonmodal, sundaram2020effects}. This feature is also noted in the vorticity contours shown in the bottom frames. When one compares the cases with and without the bump, the flow separation can be inferred as the location where vorticity has zero-contour value.

\section{Bypass Transition Caused by Free-Stream Convecting Vortex with and without the Surface Bump}

The vorticity dynamics pertaining to the test cases with and without the surface bump on a semi-infinite flat plate are
compared in this section to ascertain the cumulative role of discrete roughness and convecting vortex in the receptivity of
the boundary layer. The most optimal speed of convection, $c = 0.3$, as demonstrated previously \cite{sundaram2020effects, sengupta2018enstrophy, sengupta2019direct}, has been used for all the cases. The strength of the vortex is $\Gamma = 5$.

In prior experimental \cite{lim2004visual} and numerical \cite{sengupta2003vortex, sengupta2019direct, sengupta2020nonmodal} studies, it has been demonstrated that a free stream convecting vortex with anti-clockwise circulation creates an instability ahead of it, due to a locally imposed adverse pressure gradient in the boundary layer. The boundary conditions created by the finite core translating vortex have been described in the previous section, indicating the displacement and circulatory components of the imposed stream function on the far field boundary. It has been shown from the numerical solution of the NSE that, unsteady separation occurs at a lower value of adverse pressure gradient than that for steady separation \cite{telionis1981unsteady, kumar2026supervised, sengupta2025bifurcation}. The imposed pressure gradient at the edge of the undisturbed shear layer is indicated by the Falkner-Skan parameter, $m$, which can be far in excess of what is needed for steady separation, in the case of unsteady separation. Thus, for vortex-induced instability, it is not merely the strength of adverse pressure gradient that triggers unsteady separation, rather it is the time duration over which the adverse pressure gradient acts, which is more important \cite{sengupta2020numerical, sundaram2020effects}.

Once the equilibrium solution is obtained, the free stream vortex is started with constant convection speed ($c = 0.3$) from the initial location: ($x = -1, y = 2$). The flow is started impulsively with the initial location of the vortex being ahead of the leading edge.

\begin{figure*}
\begin{center}
\includegraphics[width=0.85\textwidth]{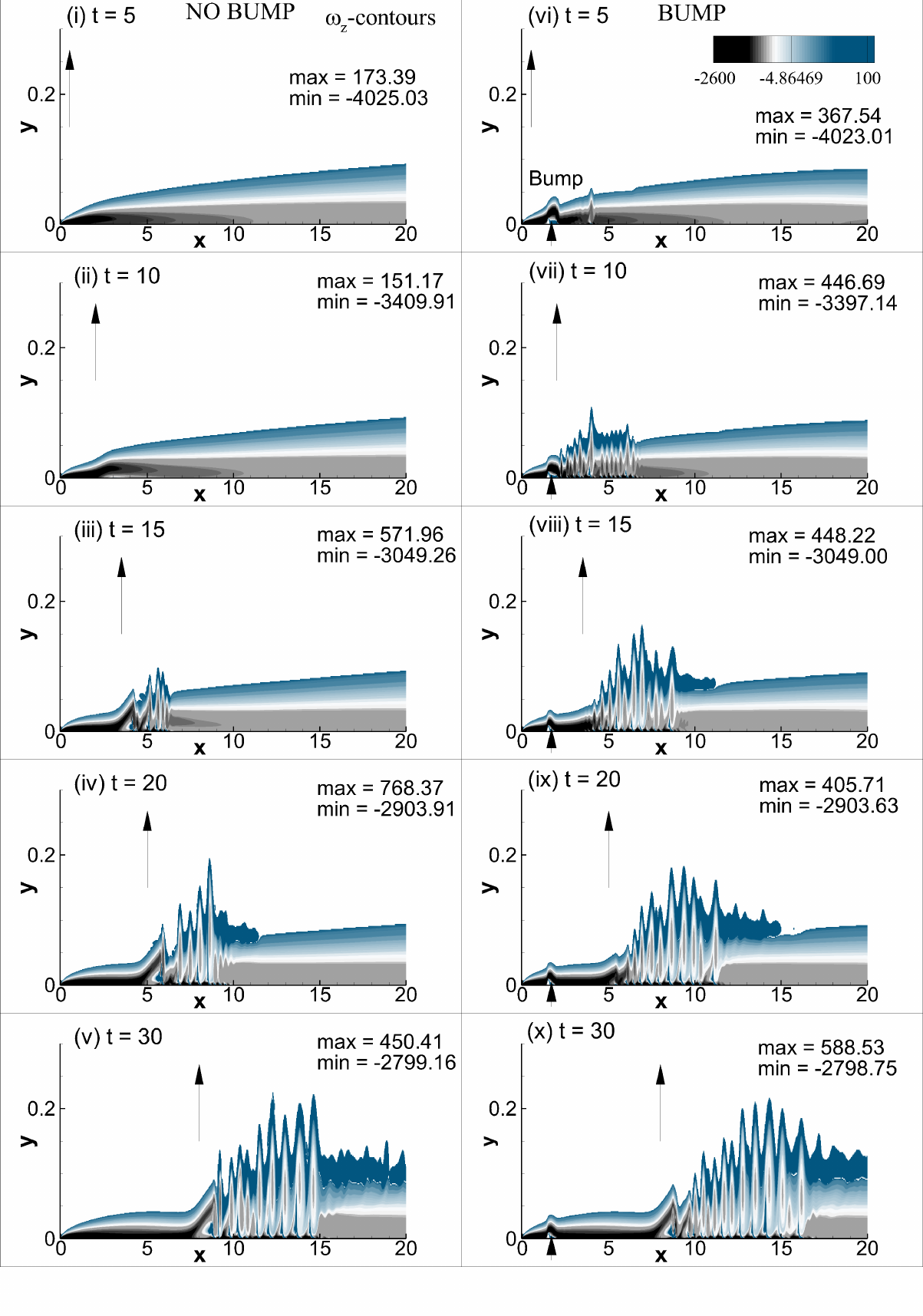}
\end{center}
\caption{Vorticity contours plotted at the indicated times for the cases with and without the bump. Arrowheads at the top show the streamwise location of the convecting vortex of strength, $\Gamma = +5$ and speed, $c =0.3$.}
\label{fig3}
\end{figure*}

In Fig. \ref{fig3}, vorticity contours are plotted for the cases with and without the bump, at selected times. The vortex with anti-clockwise circulation lifts up the boundary layer ahead of it, creating vortical structures. Upstream of the vortex, a favorable pressure gradient is created, while in the downstream direction, the induced velocity reduces with $x$, i.e. an adverse pressure gradient is created. The free stream vortex convects at a constant height and its instantaneous streamwise location is shown by an arrowhead. For both cases, there are two distinct structures in the primary stage of disturbance evolution: one that originates from the leading edge of the plate having $\omega_d < 0$ and the other originating at the wall at a downstream location having $\omega_d > 0$. This can be explained by the scouring action of the counterclockwise rotating vortex. A vortex with counterclockwise rotation lifts up the boundary layer in the downstream direction and such action is equivalent to creating an adverse pressure gradient, promoting an increase in the boundary layer thickness. This leads to the formation of nascent unsteady separation bubbles enclosing vortical structures, which eventually merge to form vortical eruptions \cite{peridier1991vortex}. At the same time, at upstream locations, a favorable pressure gradient is created, leading to simultaneous thickening and thinning of the underlying boundary layer. For the case without a bump, the disturbance packet is much more compact. As a consequence, the vortex-induced instability will create larger magnitude of wall-normal component of velocity ($v$) and the induced vortical field will have a larger positive vorticity, due to larger value of $\frac{\partial v}{\partial x}$, which can be readily noted from Fig. \ref{fig3}. The negative maximum of vorticity occurs inside the boundary layer, and is essentially due to $\frac{\partial u}{\partial y}$, and remains in the similar range. The case without a bump is representative of the unit process in FST \cite{sengupta2020numerical, sengupta2019direct} and is the building block for the physical mechanism in bypass transition, i.e. STWF. Here, no model for FST is required, with the effects governed solely by the NSE. The presence of the surface bump accelerates the onset process of bypass transition. Due to higher streamwise spreading of the packet with the surface bump, weaker vortical eruptions are noted at later times.

\begin{figure*}
\begin{center}
\includegraphics[width=0.95\textwidth]{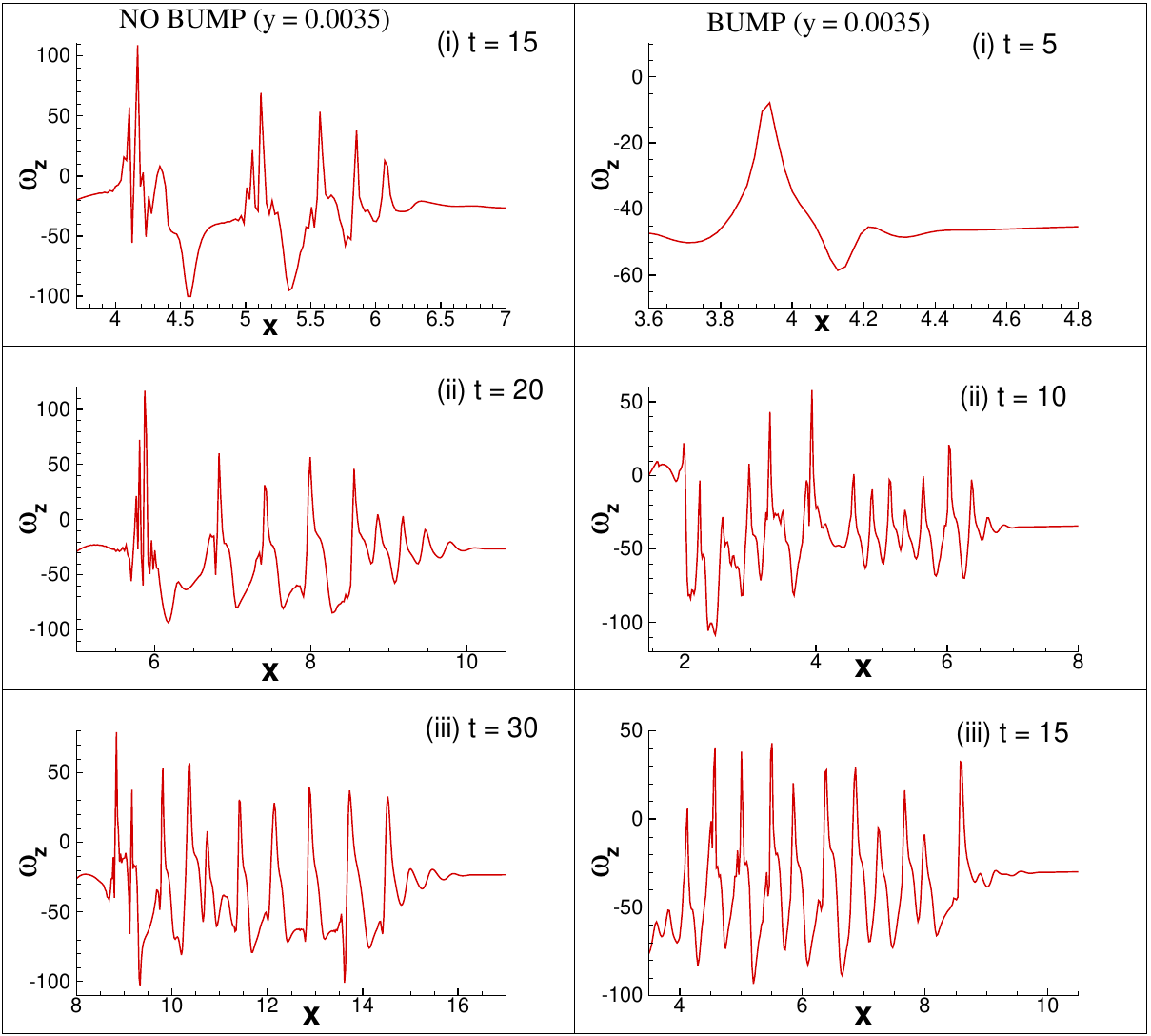}
\end{center}
\caption{Vorticity variation with streamwise distance plotted at the indicated times for the cases with and without the bump at $y= 0.00035$, for the convecting vortex moving with speed, $c =0.3$.}
\label{fig4}
\end{figure*}

The vorticity contours in Fig. \ref{fig3} show that the effect of the surface bump on the plate surface is to significantly accelerate the process of instability and transition. The specific effect of the bump can be quantified by plotting the streamwise variation of vorticity at some selected times for a specific near-wall height of $y = 0.00035$, as shown in Fig. \ref{fig4}. The presence of the bump shows the accelerated growth of disturbance, as compared to the case without bump. The time frames are matched to roughly show the correspondence of the physical events in the cases with and without bump. The streamwise affected region is wider due to the presence of the bump. Otherwise, as a consequence of the free stream convecting vortex, a STWF is created that convects downstream with time, while the presence of the bump affects the streamwise extent right from the leading edge of the flat plate. By $t =15$, one can see significant nonlinear growth of the disturbance field in the case with the bump. While one can see a very good correlation of vorticity variation with streamwise distance for heights which are very close to the wall, the variation across the boundary layer, in these cases show qualitative similarity (these variations are not shown here, as these could have some localized variations near the bump, which will be absent for the case without the bump).

\section{Application of Nonlinear Receptivity Theories: Disturbance Enstrophy Transport Equation (DETE) and Disturbance Mechanical Energy (DME)}

The problem of transition to turbulence is governed by the NSE. Thus, to pose the vortex-induced instability as a problem of receptivity, one can solve the NSE to show the connection between instability and receptivity using linear stability/receptivity theories. Similarly, nonlinear receptivity and instability have been described based on growth of disturbance mechanical energy (DME) for the instability caused by convecting vortex in the free stream \cite{sengupta2003vortex}, which is obtained by taking the divergence of the NSE. Another measure of nonlinear receptivity is in the form of the disturbance enstrophy transport equation (DETE), formulated from the vorticity transport equation by taking the curl of the NSE. This has been used to explain the vortex-induced instability \cite{sengupta2018enstrophy, sengupta2019direct} and Taylor-Green vortex problem \cite{sengupta2018non} among many other flow instabilities \cite{sengupta2019tracking, sengupta2022thermally, sengupta2023role, sengupta2024thermal}.

For fluid flows, the conservation of rotational momentum is not ensured a priori due to time-dependent inertia properties. The rotational aspect of the flow field is investigated by looking at the vorticity transport equation, obtained by taking a curl of the NSE, and rotation is quantified by the enstrophy ($\Omega_1 = \vec{\omega} \cdot \vec{\omega}$). The literature on turbulence relates enstrophy to explain the role of dissipation in determining the vorticity dynamics\cite{doering1995applied, donzis2008dissipation}. Doering and Gibbon in an attempt to study the enstrophy transport equation for 2D flow that is periodic, obtained a decaying evolution of enstrophy over the full domain, such that the exponent is given by the kinematic viscosity \cite{doering1995applied}. This is only true as a global property for a periodic problem, while for generic 3D inhomogeneous flows, the enstrophy transport equation is derived to show effects of diffusion in fluid flows \cite{kraichnan1967inertial, sengupta2013diffusion, sengupta2023multi}.

Diffusion creates rotationality for inhomogeneous flows, following which an evolution equation is developed for $\Omega_1$ as a point property \cite{sengupta2013diffusion} to explain how smaller scales are supported via the enstrophy transport process progressively. It is similar to kinetic energy describing the translational motion in fluid flows. The evolution equation shows the energy budget of the system to create and sustain rotationality. During a physical instability, the system migrates from one equilibrium state to another by redistributing energy into rotational and translational degrees of freedom. The enstrophy transport equation \cite{suman2022novel, joshi2024highly, joshi2025comparing, sengupta2023effects} is derived from the nondimensional vorticity transport equation as,

\begin{equation}
\frac{\partial \omega_i}{\partial t} + u_j \frac{\partial \omega_i}{\partial x_j} = \omega_j \frac{\partial u_i}{\partial x_j} + \frac{1}{Re} \frac{\partial^2 \omega_i}{\partial x_j \partial x_j}
\label{Eq15}
\end{equation}

\noindent where subscripts, $i,j = 1, 2\; {\rm and}\; 3$, represent Cartesian indices and follows Einstein notation. Taking a dot product of Eq. (\ref{Eq15}) with $\omega_i$, the enstrophy transport equation is obtained as follows,

\begin{dmath}
\frac{\partial \Omega_1}{\partial t} + u_j \frac{\partial \Omega_1}{\partial x_j} - 2\omega_i \omega_j \frac{\partial u_i}{\partial x_j} = \frac{1}{Re} \frac{\partial^2 \Omega_1}{\partial x_j \partial x_j} - \frac{2}{Re} \left(\frac{\partial \omega_i}{\partial x_j}\right) \left(\frac{\partial \omega_i}{\partial x_j}\right)
\label{Eq16}
\end{dmath}

It has been explained \cite{sengupta2018enstrophy} that \lq\lq{\it the third term on the left-hand side (LHS) is due to vortex stretching (corresponding to the first term on the right hand side (RHS) of Eq. (\ref{Eq15})), which is absent for 2D flows. The diffusion of $\omega_i$ gives rise to RHS terms in Eq. (\ref{Eq16}). The first term of RHS shows diffusion of $\Omega_1$ and the second term represents strictly a loss or dissipation for the transport of $\Omega_1$} \rq\rq. For 2D flows, the spanwise component of vorticity ($\omega$) is the non-zero component and in vector calculus notation the enstrophy transport equation is written as

\begin{dmath}
\frac{D \Omega_1}{Dt} = \frac{2}{Re} \left[\frac{1}{2} \nabla^2 \Omega_1 - \left|\nabla \omega \right|^2 \right]
\label{Eq17}
\end{dmath}

For inhomogeneous flows, the first term on the RHS is diffusive and can be either positive or negative. This term can create or remove rotationality according to its sign. As $\Omega_1$ is always positive, therefore a positive RHS will indicate growth of $\Omega_1$ to indicate instability. This instability gives a mechanism to create vorticity at different scales by diffusion, which is distinctly different from the creation of smaller scales by vortex stretching, as is the case for 3D flows. Thus, diffusion helps in creating new length scales for both 2D and 3D flows.

It has been shown \cite{sengupta2018non, sengupta2018enstrophy} that one can perform a regular perturbation analysis to develop a governing equation for the disturbance enstrophy transport equation by writing $\Omega_1$ as a sum of equilibrium and disturbance values as, $\Omega_1 = \Omega_m + \epsilon \Omega_d$, and linear growth of
$\Omega_d$ has been kept by retaining terms of the order of $\epsilon$. The primary variables are also similarly presented in terms of an equilibrium and a disturbance quantity with subscripts $m$ and $d$, respectively. The growth or decay rate of $\Omega_m$ is written as,

\begin{dmath*}
    \frac{D\Omega_m}{Dt} + \epsilon \frac{\partial \Omega_d}{\partial t} + \epsilon \overrightarrow{V}_m \cdot \nabla \Omega_d +
    \epsilon \overrightarrow{V}_d \cdot \nabla \Omega_m + \epsilon^2 \overrightarrow{V}_d \cdot \nabla \Omega_d = RHS
\end{dmath*}

Thus, one can obtain an evolution equation for the growth/decay rate of $\Omega_d$, which is referred to as the DETE in the literature. It is given by,

\begin{dmath}
\label{dete}
\nonumber
\frac{D\Omega_d}{Dt} = \biggl[ \biggl\{ \omega_{im} \omega_{jm} \frac{\partial u_{id}}{\partial x_j} + \omega_{im} \omega_{jd} \frac{\partial u_{im}}{\partial x_j} + \omega_{id} \omega_{jm} \frac{\partial u_{im}}{\partial x_j}\biggr\} \\
+ \frac{1}{Re} \frac{{\partial}^2\Omega_d}{\partial x_j \partial x_j} -\frac{2}{Re} \biggl (\frac{\partial \omega_{im}}{\partial x_j} \biggr) \biggl( \frac{\partial \omega_{id}}{\partial x_j} \biggr) \biggr]
\end{dmath}

The vortex stretching will cause growth/ decay of $\Omega_d$ given by the terms inside the curly brackets in Eq. \eqref{dete}. Such terms are absent for 2D disturbance flows. The last term on the RHS can also contribute to growth of disturbance field. Also, $\Omega_d$ can be either positive or negative, as this is given as ($2\overrightarrow{\omega}_m \cdot \overrightarrow{\omega}_d$). Thus, instability for $\Omega_d$ can arise for both positive and negative $\Omega_d$. For $\Omega_d > 0$, instability will be caused for $D\Omega_d/ Dt > 0$. Similarly, for $\Omega_d < 0$, its amplitude will grow, when $D\Omega_d/ Dt < 0$.

In an attempt to explain turbulence, Landahl and Mollo-Christensen\cite{landahl1992turbulence} emphasized that one should consider growth of total mechanical energy and not just the turbulent kinetic energy. This prompted Sengupta {\it et al.}\cite{sengupta2003vortex, sengupta2019direct} to develop the equation for the mechanical energy from the NSE, by writing it in rotational form as,

\begin{dmath}
\frac{\partial \overrightarrow{V}}{\partial{t}} -\overrightarrow{V} \times \overrightarrow{\omega} = - \nabla {\biggl(\frac {p}{\rho}+\frac{|\overrightarrow{V}|^2}{2}\biggr)}  - \nu (\nabla \times \overrightarrow{\omega})
\label{DME2}
\end{dmath}

Defining the total mechanical energy as
$$E = \frac{p}{\rho} + \frac{|\overrightarrow{V}|^2}{2}$$

\noindent one can take the divergence of the NSE and use the incompressibility condition to get the Poisson equation for $E$ as,

\begin{equation}
 \nabla^2 E = \nabla \cdot (\overrightarrow{V} \times \overrightarrow{\omega})
\label{DME4}
\end{equation}

Using the vector identity,
$$ \nabla \cdot (\overrightarrow{V} \times \overrightarrow{\omega}) = \overrightarrow{\omega} \cdot (\nabla \times \overrightarrow{V}) - \overrightarrow{V} \cdot (\nabla \times \overrightarrow{\omega})$$

\noindent one finally gets the governing equation for the total mechanical energy as

\begin{dmath}
\nabla^2 E = \overrightarrow{\omega}\cdot \overrightarrow{\omega} -\overrightarrow{V} \cdot (\nabla \times \overrightarrow{\omega})
\label{DME5}
\end{dmath}

This derivation of the boundary value problem for $E$ is from the NSE without any assumption, other than that of incompressibility. The contribution to $E$ comes from the convective acceleration and pressure gradient term, while the local acceleration and the viscous actions are not directly present. The distribution of $E$ is due to forcing by the enstrophy. Rotationality caused by viscous diffusion or other sources will affect $E$, through its forcing by enstrophy.

Without solving Eq. (\ref{DME5}), one interprets the solution, by the sign of the RHS of the Poisson equation \cite{sommerfeld1949partial}. It shows a source or a sink of $E$, with a negative/positive sign, respectively. Thus, it may appear that the enstrophy stabilizes $E$. But, the growth of $E$ by instability of the equilibrium state has been studied by noting, $E = E_m + \epsilon E_d$, with subscripts $m$ and $d$ indicating the equilibrium and disturbance components, respectively. For the growth of small perturbations, the parameter $\epsilon$ represents a small value, and one can split all quantities appearing in Eq. (\ref{DME5}). The equation for disturbance energy, $E_d$, is obtained by subtracting the equation for $E_m$ from Eq. \eqref{DME5} as,

\begin{dmath}
\nabla^2 E_d = \Omega_d + \epsilon \overrightarrow{\omega}_d \cdot \overrightarrow{\omega}_d - \overrightarrow{V}_m \cdot (\nabla  \times \overrightarrow{\omega}_d) - \overrightarrow{V}_d \cdot (\nabla \times \overrightarrow{\omega}_m) - \epsilon \overrightarrow{V}_d \cdot (\nabla \times \overrightarrow{\omega}_d)
\label{DME6}
\end{dmath}

We have described the criteria of linear and nonlinear instabilities with the help of DETE and DME equations. Next, the conditions of instability with these nonlinear theories are investigated for vortex-induced instability affected by the presence or absence of the 2D surface roughness.

\begin{figure*}
\begin{center}
\includegraphics[width=\textwidth]{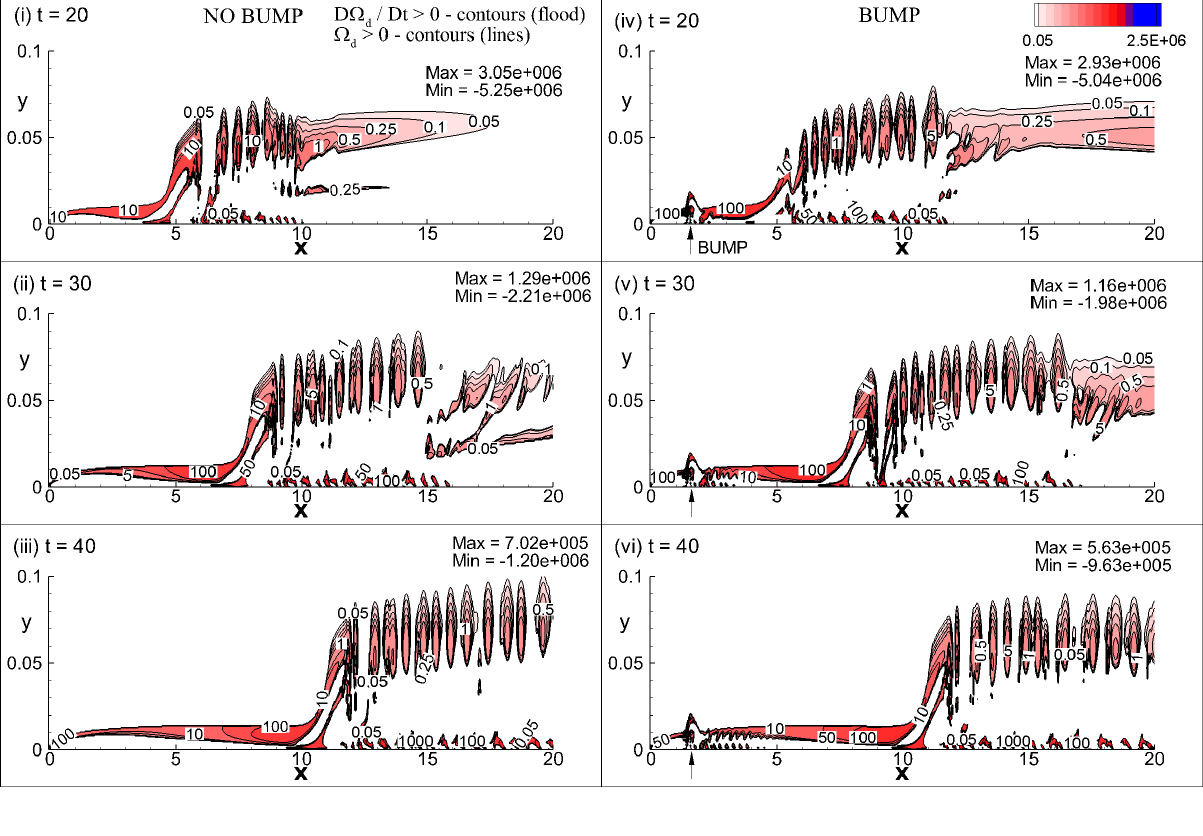}
\end{center}
\caption{The growth rates of $\Omega_d$ are compared for the vortex-induced instability problem, with and without the 2D surface bump, shown for positive $\Omega_d$ at the indicated time instants for the case of $c =0.3$.}
\label{fig5}
\end{figure*}

\begin{figure*}
\begin{center}
\includegraphics[width=\textwidth]{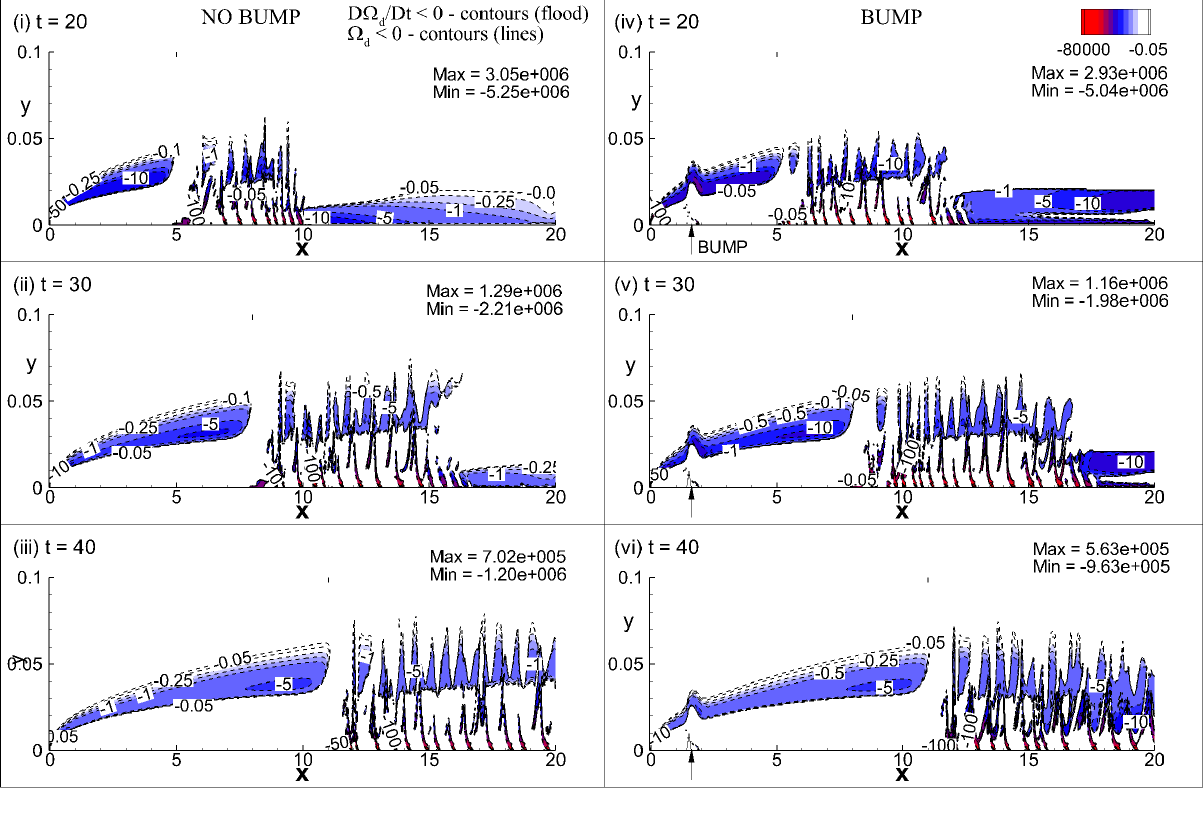}
\end{center}
\caption{The growth rate of $\Omega_d$ are compared for the vortex-induced instability problem, with and without the 2D
surface roughness, shown for negative $\Omega_d$ at the indicated time instants for the case of $c = 0.3$.}
\label{fig6}
\end{figure*}

In Figs. \ref{fig5} and \ref{fig6}, the growth rate contours of $D\Omega_d/ Dt$ are plotted in the domain, for $\Omega_d > 0$ and for $\Omega_d < 0$, respectively, for the free stream vortex convecting with speed, $c = 0.3$. The ensuing vortex-induced instability with and without the surface roughness is compared. The unstable regions are shown by the flooded contours in the various frames. The regions of instability are qualitatively different for positive and negative values of $\Omega_d$. These regions are complementary, i.e. where one notices instability for $\Omega_d < 0$, the same region indicates no instability for the case of $\Omega_d > 0$ and vice versa. For the positive values, disturbances originate outside the shear layer, and these are affected by the vortex-induced wall-normal eruptions, as shown at $t = 20$ for cases with and without the bump. As a consequence of these ejections starting from the wall, one notices the formation of oblong vortical structures. In the downstream direction, these vortical structures are followed by weaker streaky disturbance growth region \cite{saric2002boundary, sundaram2020effects}. For the case with bump, this streaky region is more elongated. However by $t = 30$, the growth rates weaken for both cases (as noted by the maximum value), but the oblong vortical structures are more numerous and the horizontal structure is punctuated by the formation of inclined growth regions. It is noted that the presence of bump reduces growth in all the frames. At $t = 40$, the maximum growth rates come down even further, and one notices convection of vortical structures. One significant difference for the case with bump is the creation of unsteady separation on the wall, immediately downstream of it and this region is elongated with time. In Fig. \ref{fig6} at $t = 20$, larger growth rates near the wall are noted. This can be interpreted as an indicator for predominance of wall-mode compared to the free stream-mode for $\Omega_d < 0$. With time, the wall-modes are shown to eject vortical structures in the wall-normal direction, as seen at $t = 30$. By later time, large growth rates are seen to be confined inside the shear layer, while the outer disturbances for downstream $x$-stations have smaller growth rates. In Figs. \ref{fig5} and \ref{fig6}, the growth rate contours are seen to create multiple decks.

\section{Three-Dimensional DNS of Vortex-Induced Instability Transition Due to Surface Bump on a Semi-Infinite Flat Plate}

In this section, we extend our study on the effect of introducing a Gaussian bump in flow past a semi-infinite flat plate subjected to a free stream convecting vortex by computing the 3D flow field. In the absence of a bump, the disturbance field remains primarily 2D due to the nature of excitation. This is not the case in the presence of  the bump, which triggers early transition and inception of three-dimensionality in the flow.

\begin{figure*}
\begin{center}
\includegraphics[width=0.95\textwidth]{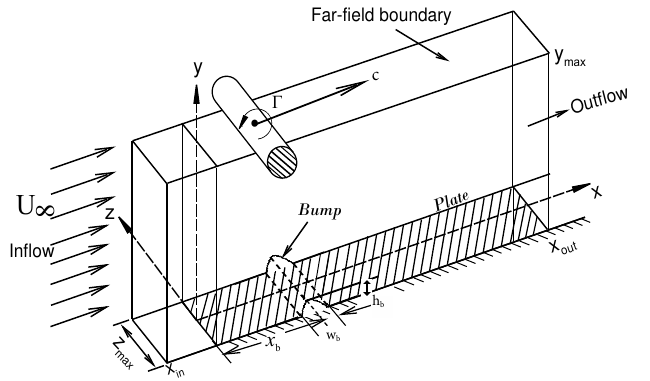}
\end{center}
\caption{Schematic diagram of the 3D receptivity problem for the boundary layer forming over a semi-infinite flat plate, with a Gaussian bump.}
\label{fig7}
\end{figure*}

In Fig. \ref{fig7}, a schematic for the 3D receptivity case for flow past a semi-infinite flat plate to free stream convecting vortex excitation is shown. There is a Gaussian bump placed near the leading edge, as in Fig. \ref{fig1} for the 2D flow. The leading edge of the plate is kept inside the computational domain, with the origin placed at the mid-point of the leading edge of the plate. The domain is given as $x_{in} \le x \le x_{out}$ in the streamwise direction with $x_{in} =-0.05$, $x_{out} =20$; $0 \le y \le y_{max}$ in the wall-normal direction with $y_{max} = 0.75$, and $z_{max}/2 \le z \le z_{max}/2$ with $z_{max} = 1.6$ in the spanwise direction. The notations and parameters used are the same as in Sengupta {\it et al.} \cite{sengupta2020numerical, sengupta2019direct}.

The simulations are performed in the uniformly spaced transform $(\xi, \eta, \zeta)$-plane, with $x=x(\xi)$, $y=y(\eta)$ and $z=z(\zeta)$. The rotational form of the $(\protect\overrightarrow{V}, \protect\overrightarrow{\omega})$-formulation of the incompressible NSE is used here \cite{sengupta2013high} which is given as

\begin{equation}
\frac{\partial \omega_{\xi}}{\partial t}+
\biggl( \frac{1}{h_2} \frac{\partial H_{\zeta}}{\partial \eta} - \frac{1}{h_3} \frac{\partial H_{\eta}}{\partial \zeta}\biggl) = 0
\label{6eq1}
\end{equation}
\begin{equation}
\frac{\partial \omega_{\eta}}{\partial t}+
\biggl( \frac{1}{h_3} \frac{\partial H_{\xi}}{\partial \zeta} - \frac{1}{h_1} \frac{\partial H_{\zeta}}{\partial \xi}\biggl) = 0
\label{6eq2}
\end{equation}
\begin{equation}
\frac{\partial \omega_{\zeta}}{\partial t}+
\biggl( \frac{1}{h_1} \frac{\partial H_{\eta}}{\partial \xi} - \frac{1}{h_2} \frac{\partial H_{\xi}}{\partial \eta}\biggl) = 0
\label{6eq3}
\end{equation}

Here, the vorticity and velocity vectors are given as $\protect\overrightarrow{\omega}=(\omega_{\xi}, \omega_{\eta}, \omega_{\zeta})$ and $\protect\overrightarrow{V}=(u, v, w)$. The grid transformation is given by the scale factors $h_1$, $h_2$ and $h_3$ given as, $h_1=\frac{\partial x}{\partial \xi}$, $h_2=\frac{\partial y}{\partial \eta}$ and $h_3=\frac{\partial z}{\partial \zeta}$. In Eqs.~\eqref{6eq1} to \eqref{6eq3}, the terms $H_{\xi}$, $H_{\eta}$ and $H_{\zeta}$ are given as

\begin{equation}
H_{\xi} = (w \omega_{\eta} - v \omega_{\zeta}) + \frac{1}{Re_L}
\biggl( \frac{1}{h_2} \frac{\partial \omega_{\zeta}}{\partial \eta} - \frac{1}{h_3} \frac{\partial \omega_{\eta}}{\partial \zeta}\biggl)
\label{6eq4}
\end{equation}
\begin{equation}
H_{\eta} = (u \omega_{\zeta} - w \omega_{\xi}) + \frac{1}{Re_L}
\biggl( \frac{1}{h_3} \frac{\partial \omega_{\xi}}{\partial \zeta} - \frac{1}{h_1} \frac{\partial \omega_{\zeta}}{\partial \xi}\biggl)
\label{6eq5}
\end{equation}
\begin{equation}
H_{\zeta} = (v \omega_{\xi} - u \omega_{\eta}) + \frac{1}{Re_L}
\biggl( \frac{1}{h_1} \frac{\partial \omega_{\eta}}{\partial \xi} - \frac{1}{h_2} \frac{\partial \omega_{\xi}}{\partial \eta}\biggl)
\label{6eq6}
\end{equation}

In this $(\protect\overrightarrow{V},\protect\overrightarrow{\omega})$-formulation, the velocity is obtained from the Poisson equations $\nabla^2{\protect\overrightarrow{V}} = -\nabla\times \protect\overrightarrow{\omega}$, those are given in the transformed plane as

\begin{equation}
\nabla^2_{\xi\eta\zeta} u = \biggl( h_1 h_2 \frac{\partial \omega_{\eta}}{\partial \zeta} - h_3 h_1 \frac{\partial \omega_{\zeta}}{\partial \eta} \biggl)
\label{6eq7}
\end{equation}
\begin{equation}
\nabla^2_{\xi\eta\zeta} v = \biggl( h_2 h_3 \frac{\partial \omega_{\zeta}}{\partial \xi} - h_1 h_2 \frac{\partial \omega_{\xi}}{\partial \zeta} \biggl)
\label{6eq8}
\end{equation}
\begin{equation}
\nabla^2_{\xi\eta\zeta} w = \biggl( h_3 h_1 \frac{\partial \omega_{\xi}}{\partial \eta} - h_2 h_3 \frac{\partial \omega_{\eta}}{\partial \xi} \biggl)
\label{6eq9}
\end{equation}

\noindent where, the Laplacian operator $\nabla^2_{\xi\eta\zeta}$ is given as

\begin{dmath}
     h_1h_2h_3\;\nabla^2_{\xi\eta\zeta} = \frac{\partial}{\partial \xi } \biggl(\frac{h_2 h_3}{h_1} \frac{\partial }{\partial \xi } \biggl) + \frac{\partial}{\partial \eta} \biggl(\frac{h_3 h_1}{h_2} \frac{\partial }{\partial \eta} \biggl) +
    \frac{\partial}{\partial \zeta} \biggl(\frac{h_1 h_2}{h_3} \frac{\partial }{\partial \zeta} \biggl)
\end{dmath}

The Reynolds number based on $L$ is $Re_L=U_{\infty}L/\nu=10^5$, for the results presented here. The strength of the free stream convecting vortex is given by $\Gamma = 2$ and it is convecting downstream from left to right with speed, $c = 0.3$.

Following the procedure used for 2D problem in Sec. II, here also, the boundary conditions at inflow and at the far-field are found by including induced velocity and vorticity, obtained from Biot-Savart law. Periodic condition for all the three components of velocity and three components of vorticity is used in the spanwise direction. At the wall, no-slip conditions on $u$, $v$ and $w$ are used. The boundary conditions for the vorticity components on the plate surface are obtained from

\begin{equation}
\omega_{\eta} = 0
\end{equation}
\begin{equation}
\omega_{\xi} = \biggl(\frac{\partial w}{\partial y}-\frac{\partial v_w}{\partial z} \biggl) \;\;\; {\rm and} \;\;\omega_{\zeta} = \biggl( \frac{\partial v_w}{\partial x} - \frac{\partial u}{\partial y} \biggl)
\label{6eq14}
\end{equation}

Since the sharp leading edge of the flat plate is the locus of stagnation points for the flow, ahead of the leading edge for the bottom plane, symmetry conditions are used for all the six variables,
\begin{equation}
\frac{\partial u}{\partial \eta} = v = \frac{\partial w}{\partial \eta} = \omega_{\xi} = \frac{\partial \omega_{\eta}}{\partial \eta} = \omega_{\zeta} = 0
\label{6eq15}
\end{equation}

At the outflow, the convective boundary conditions are applied on the variables $u$, $\omega_{\eta}$ and $\omega_{\zeta}$,
\begin{equation}
\frac{\partial u}{\partial t} + U_c \frac{\partial u}{\partial x} = 0
\label{6eq16}
\end{equation}
\begin{equation}
\frac{\partial \omega_{\eta}}{\partial t} + U_c \frac{\partial \omega_{\eta}}{\partial x} = 0
\label{6eq17}
\end{equation}
\begin{equation}
\frac{\partial \omega_{\zeta}}{\partial t} + U_c \frac{\partial \omega_{\zeta}}{\partial x} = 0
\label{6eq18}
\end{equation}

The boundary condition for $\omega_{\xi}$ at the outflow boundary is given by the solenoidality condition of vorticity as
\begin{equation}
\frac{\partial \omega_{\xi}}{\partial x} = -\biggl( \frac{\partial \omega_{\eta}}{\partial y} + \frac{\partial \omega_{\zeta}}{\partial z} \biggl)
\label{6eq19_N}
\end{equation}

\begin{figure*}
\begin{center}
\includegraphics[width=0.95\textwidth]{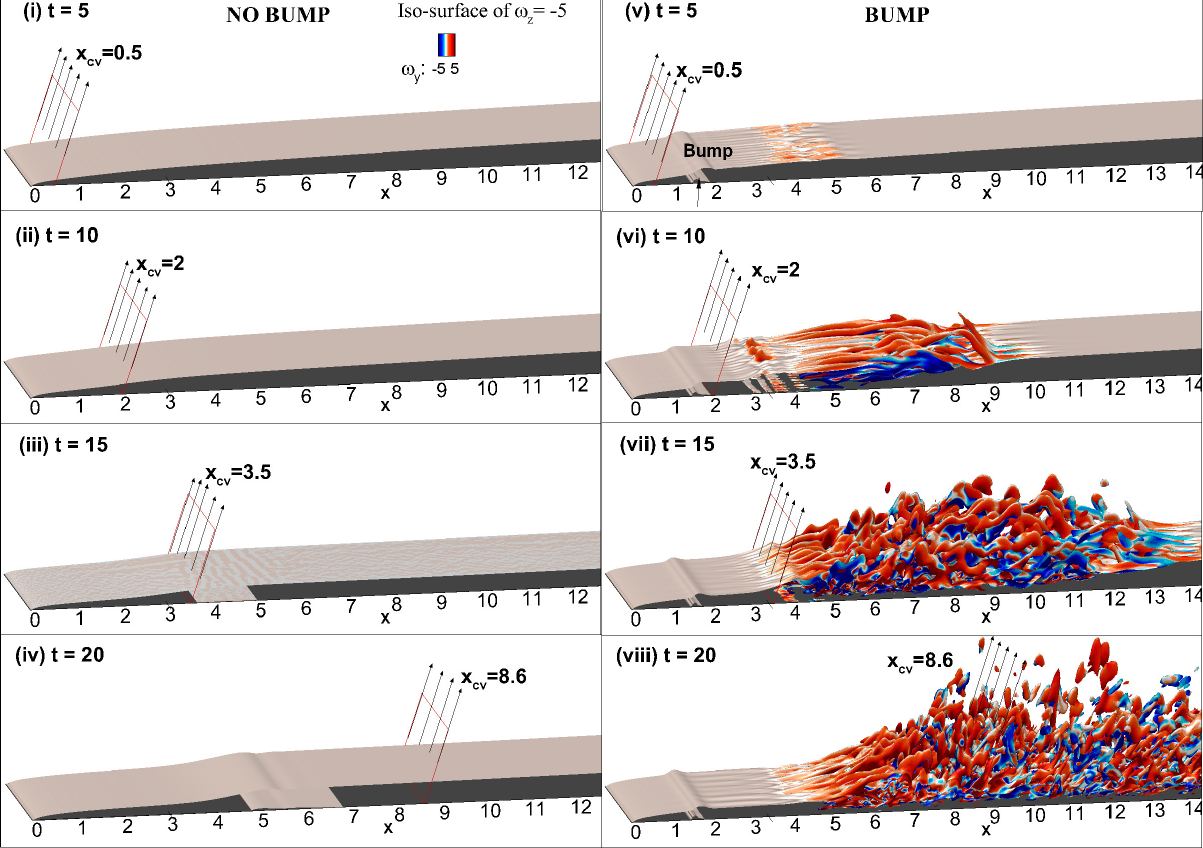}
\end{center}
\caption{Comparison of spanwise vorticity iso-contour value of $\omega_{\zeta} = -5$ at the indicated times for the cases with and without the bump. The contours are colored with $\omega_{\eta}$ in the range $\pm 5$. Arrowheads show the streamwise location of the convecting positive vortex (of strength $\Gamma=2$) moving with speed, $c =0.3$.}
\label{fig8}
\end{figure*}

\noindent The boundary conditions for the $v$- and $w$-components of velocity at the outflow boundary are
derived from the definition of the vorticity components $\omega_{\eta}$ and $\omega_{\zeta}$ as
\begin{equation}
\frac{\partial v}{\partial x} = \frac{\partial u}{\partial y} + \omega_{\zeta}
\label{6eq19}
\end{equation}
\begin{equation}
\frac{\partial w}{\partial x} = \frac{\partial u}{\partial y} - \omega_{\eta}
\label{6eq20}
\end{equation}

With the 2D equilibrium flow obtained for the cases with and without the bump, the 3D equilibrium flows are obtained by specifying $u$, $v$ and $\omega_{\zeta}$ variables at all the discrete spanwise stations, while prescribing other variables, $w$, $\omega_{\xi}$ and $\omega_{\eta}$ as zero at all locations. Subsequently, free stream excitation by the convecting vortex is initiated. The numerical methods and other parameters are as given in Sengupta {\it et al.} \cite{sengupta2020numerical, sengupta2019direct}.

In Fig. \ref{fig8}, iso-contours of spanwise vorticity ($\omega_{\zeta} = -5$) are compared for cases without bump (left) and with bump (right) at the indicated times. The contours are colored with $\omega_{\eta}$ in the range $\pm 5$. Comparing this 3D flow field with the 2D flow case shown in Fig. \ref{fig3}, one notes that three-dimensionality accelerates the transition process significantly in the presence of the bump for the 3D disturbance field. Even at $t =5$, significant 3D nature of the flow is noted in the presence of the bump. By $t =20$, one can see small scale 3D disturbance field for the case with the bump, whereas for the case without the bump the flow remains 2D for the shown time frames. Thus, the discrete roughness element changes the transition pattern drastically, by accelerating the process as compared to the 2D flow field computed in the previous section. The inherently 2D nature of the disturbance field in the absence of roughness element is compromised in the presence of roughness, thereby necessitating a full 3D DNS for the test case with roughness.

\begin{figure*}
\includegraphics[width=0.95\textwidth]{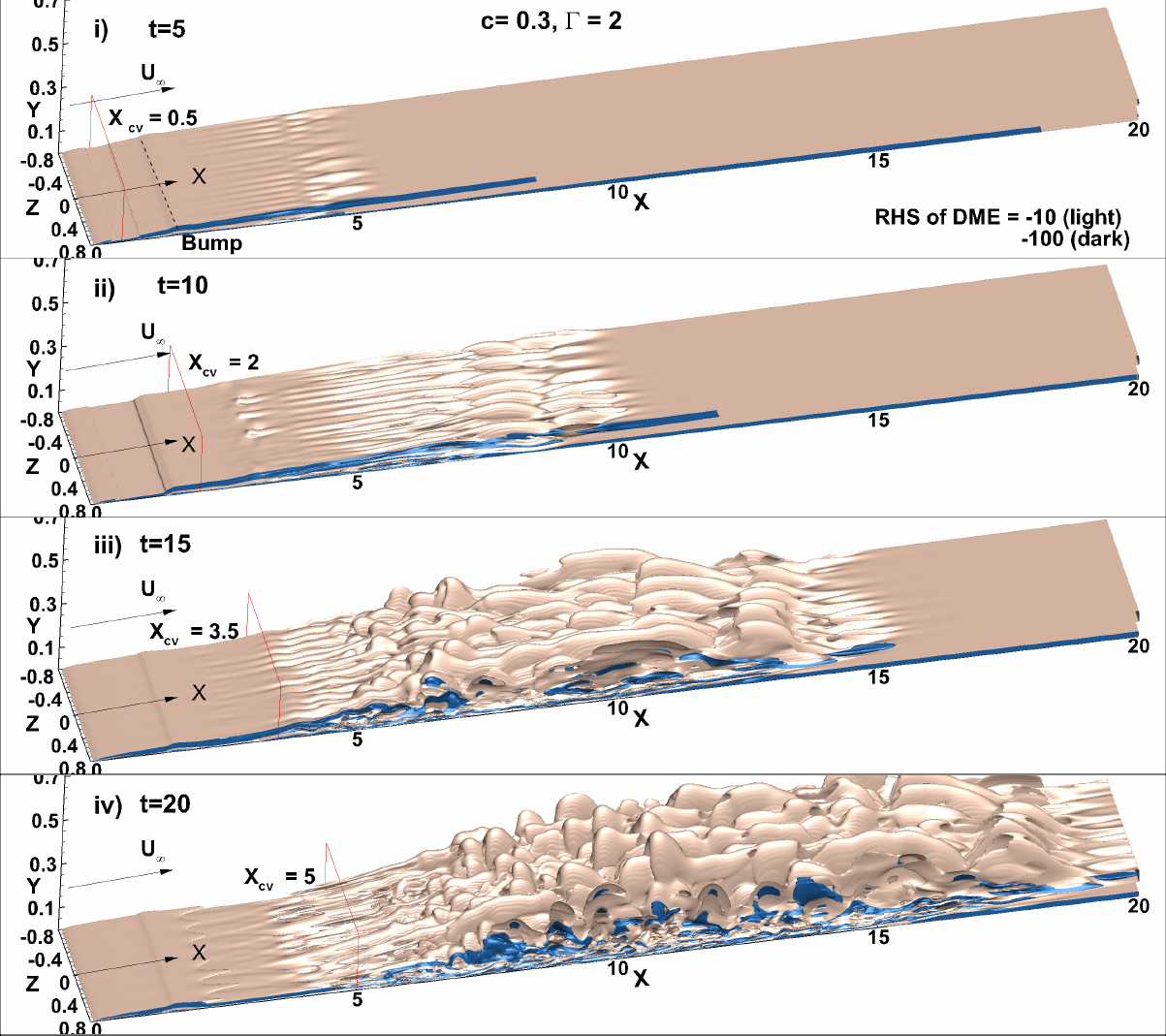}
\caption{The contour plot in the physical plane for the forcing term of the Poisson equation for the disturbance mechanical energy (Eq. \eqref{DME6}) for the free stream excitation of convecting vortex for the case with the bump.}
\label{fig9}
\end{figure*}

In Fig. \ref{fig9}, the forcing term for the DME, as given by Eq. \eqref{DME6} is shown for free stream convecting vortex in the presence of 2D surface bump, computed for 3D disturbance field. Only two contour values, indicating \lq source of DME' with magnitude of -10 (light) and -100 (dark), are shown for the indicated times. At $t = 5$, one notes the dark contours at the edge of the spanwise boundaries, which have been treated as periodic boundaries. However, two clusters of lighter structure are noted across the full span in the form of STWFs, which start from the location of the bump. This type of formation of STWF as the precursor of transition has been noted for 3D DNS of vortex-induced instability without surface roughness \cite{sengupta2019direct} at significantly later times. At $t = 20$, the full domain is noted to be affected by the DME structures starting from the streamwise location of the bump to the outflow of the computational domain. One however, notes that the disturbance structures indicated by this theory of DME are well rounded and larger, compared to the vortical structures noted in Fig. \ref{fig8}. It should be noted here that, in the presence of the bump, the STWF is an indicator of the nonmodal nature of the disturbance field, which is consistent with prior roughness-induced transition studies \cite{rizzetta2007direct, de2013laminar, reshotko2004role, suryanarayanan2019roughness}. In general, the cumulative response to the surface bump and convecting vortex can be separated into four stages: (i) generation of vortical disturbances at the roughness ($t = 5$), (ii) a spatio-temporal amplification of the disturbance field ($t = 10$), (iii) emergence and amplification of the STWF marked by sequence of unsteady separation bubbles ($t = 15$), and (iv) emergence of turbulent flow due to merging of the  unsteady separation bubbles to turbulent spots ($t = 20$).

\begin{figure*}
\begin{center}
\includegraphics[width=\textwidth]{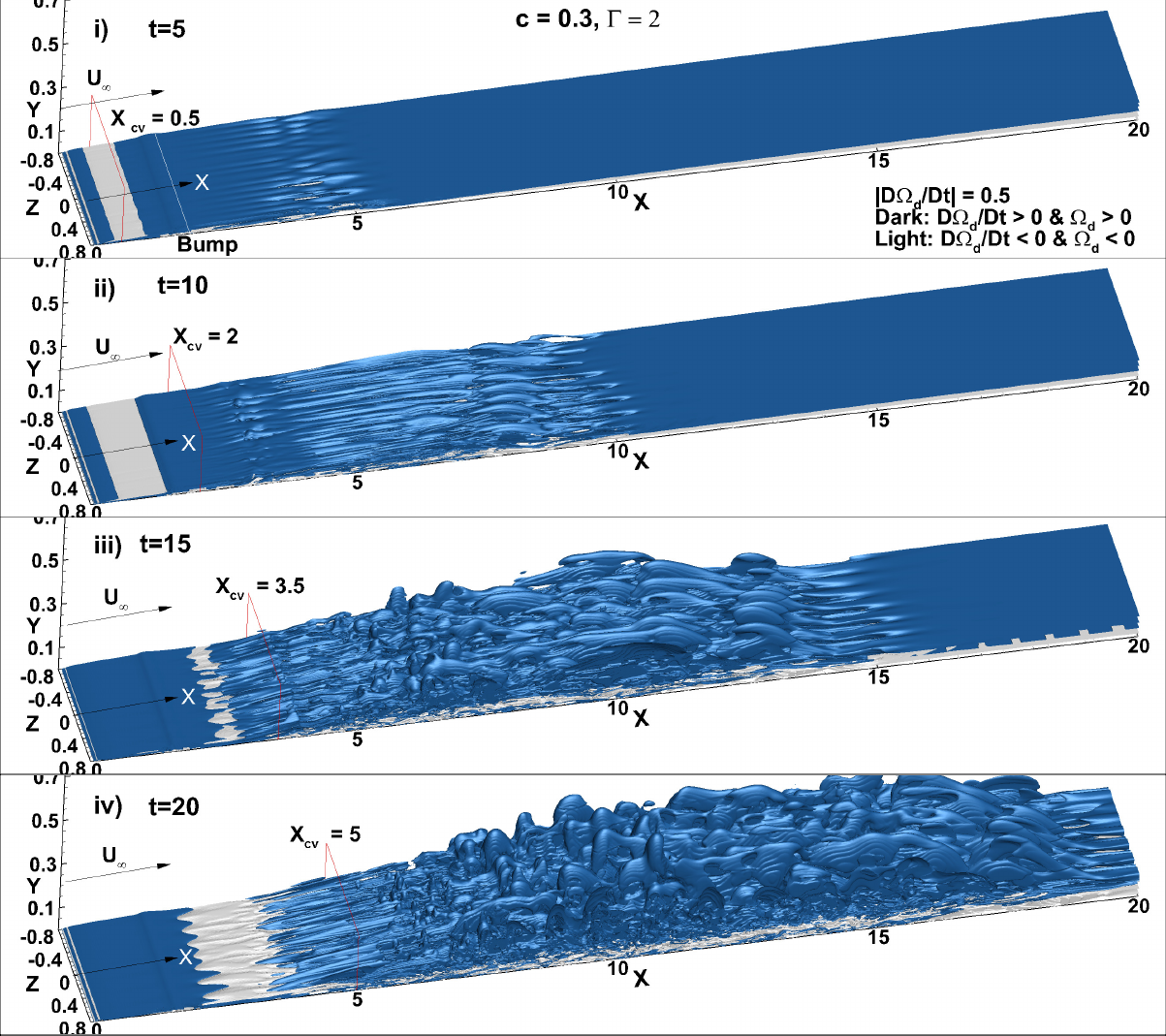}
\end{center}
\caption{The growth rates of $\Omega_d$ are compared for the vortex-induced instability problem with the 2D surface bump shown for positive and negative $\Omega_d$ at the indicated time instants for the case of $c =0.3$.}
\label{fig10}
\end{figure*}

Coherent structures detected by DME equation more diffused, compared to those obtained in direct simulation of the NSE. Different coherent structure tracking methods have been studied recently in the literature \cite{sengupta2019direct}. It has been noted that despite not making any assumptions in developing the DME equation for incompressible flows, the solenoidality condition of the velocity vector removes the unsteady and viscous diffusion terms. In comparison, the tracking method based on DETE reproduces the vortical structures quite clearly. This has already been shown for different 2D routes of transition in Sec. IV. Next, we report the structures tracked by the DETE method for the 3D receptivity problem sketched in Fig. \ref{fig7}. As noted in Sec. IV, disturbance enstrophy, $\Omega_d$, can take both signs and hence the conditions of instability stated following Eq. \eqref{dete} will also hold for this 3D receptivity problem. In Fig. \ref{fig10}, both conditions of instability with $\Omega_d > 0$ and $\Omega_d < 0$ are shown as dark and light contours, respectively. The bump location is indicated by the white line in the top frame. At $t = 5$, one notes instability for $\Omega_d < 0$ in a patch upstream of the surface bump. This is followed by the dark contours with $\Omega_d > 0$, which corresponds to the regions where $\omega_d$ is of the same sign as the vorticity for the equilibrium state given by $\omega_m$. The presence of instabilities with $\Omega_d < 0$ in the near-wall regions and with $\Omega_d > 0$ in the far stream regions is consistent with prior findings for the vortex-induced instability in the absence of roughness \cite{sengupta2019direct, sengupta2020numerical, pal2026direct}. Here, only the growth rate given by $|\frac{D\Omega_d}{Dt}| = 0.5$ is shown in the figure, for the sake of clarity. If one were to chose smaller values of the growth rates, then the disturbance enstrophy iso-contours would capture smaller scale structures, similar to the vortical structures shown on right column of Fig. \ref{fig8}, for the case with the bump. Thus, DETE tracking method is found to be the most effective tool for indicating receptivity/ instability and capturing coherent structures during the nascent and transient stages of disturbance growth caused by the localized discrete surface bump over the semi-infinite flat plate.

\section{Summary and Conclusions}
Receptivity of flow over a semi-infinite flat plate to a deterministic free stream excitation in the form of a free stream convecting vortex of constant strength and speed is investigated in the presence of a surface bump, as shown in Figs. \ref{fig1} and \ref{fig7}. The aim of the present investigation is to determine whether the nonmodal nonlinear route of transition to turbulence established for the vortex-induced instability problem \cite{sengupta2020nonmodal, sundaram2020effects, sengupta2025compressible} holds when a surface roughness is imposed in the flow. Prior studies on roughness-induced transition \cite{suryanarayanan2019roughness, rizzetta2007direct, de2013laminar} have demonstrated that the bypass transition is  nonmodal in nature. Here, we show that this is also the case for the cumulative response to free stream excitation and surface bump through the vorticity contours in Figs. \ref{fig3} and \ref{fig8}. The presence of roughness element triggers earlier transition which is quantified through the streamwise variation of vorticity in Fig. \ref{fig4}. The roughness element shows the accelerated growth of disturbance in a spatio-temporal sense, as compared to the case without bump. The nonlinear theory based on DETE is applied for the 2D flow in Figs. \ref{fig5} and \ref{fig6} which show that the regions of instability are qualitatively different for positive and negative values of $\Omega_d$. The regions are complementary, i.e. where one notices instability for $\Omega_d < 0$, the same region indicates no instability for the case of $\Omega_d > 0$ and vice versa.

The iso-contours of spanwise vorticity are compared for cases with and without the surface bump in Fig. \ref{fig8} for the 3D DNS. Comparing this 3D flow field with the 2D flow case shown in Fig. \ref{fig3}, it can be inferred that three-dimensionality accelerates the transition process significantly in the presence of the bump. The inherently 2D nature of the disturbance field in the absence of roughness element is compromised in the presence of roughness, thereby necessitating a full 3D DNS for the test case with roughness. The source of the DME equation is plotted for the 3D DNS in Fig. \ref{fig9}, which demonstrates that the nonmodal nonlinear route for transition to turbulence is followed for both cases with and without the bump. The case with the surface bump and convecting vortex can be separated into four stages: (i) generation of vortical disturbances at the roughness, (ii) a spatio-temporal amplification of the disturbance field, (iii) emergence and amplification of the STWF marked by sequence of unsteady separation bubbles, and (iv) emergence of turbulent flow due to merging of the  unsteady separation bubbles to turbulent spots. The application of the nonlinear receptivity theory of DETE in Fig. \ref{fig10} shows the presence of instabilities with $\Omega_d < 0$ in the near-wall regions and with $\Omega_d > 0$ in the far stream regions. This is consistent with prior studies for the vortex-induced instability in the absence of roughness \cite{sengupta2024separation, sengupta2019tracking, sengupta2020numerical}.

One of the difficulties with transition prediction is the existence of the multiple routes \cite{saric2002boundary, morkovin1993bypass} and the different mechanisms associated with each route. The present analysis of the vortex-induced instability problem over a semi-infinite flat plate with surface bump suggests that some features may be common to other routes to transition, such as dominance of the nonmodal nature of transition. Future work would involve application of the understanding achieved for this vortex- and roughness-induced transition in targeted flow control such as introducing a second roughness element downstream of the given roughness element, such that it disrupts the emergence of the STWF, thereby suppressing transition.

\bibliography{apssamp}



\end{document}